\documentclass[conference,10pt]{IEEEtran}
\usepackage{amsthm}
\usepackage{amsmath}
\usepackage{amssymb}
\usepackage{algpseudocode}
\usepackage{bm}
\usepackage{booktabs}
\usepackage{url}
\usepackage{cite}
\usepackage{color}
\usepackage{float}
\usepackage{graphicx}
\usepackage{mathrsfs}
\usepackage{multicol}
\usepackage{multirow}
\usepackage{stfloats}

\makeatletter

\floatstyle{ruled}
\newfloat{algorithm}{tbp}{loa}
\providecommand{\algorithmname}{Algorithm}
\floatname{algorithm}{\protect\algorithmname}
\usepackage{tikz}
\usetikzlibrary{calc}

\theoremstyle{plain}

\theoremstyle{plain}

\theoremstyle{plain}

\theoremstyle{plain}

\IEEEoverridecommandlockouts
\author{\IEEEauthorblockN{Wenjie Liu and
Panos Papadimitratos}
\IEEEauthorblockA{\textit{Networked Systems Security Group}\\
\textit{KTH Royal Institute of Technology}\\
Stockholm, Sweden\\
\textit{wenjieli@kth.se},
\textit{papadim@kth.se}}
\thanks{This work was supported in part by the SSF SURPRISE cybersecurity project, the Security Link strategic research center, the Karl Engver's Foundation, and the China Scholarship Council. The computations were enabled by resources provided by the National Academic Infrastructure for Supercomputing in Sweden (NAISS), partially funded by the Swedish Research Council through grant agreement no. 2022-06725. We would also like to acknowledge the work of the organizers of Jammertest 2024.}
}


\usepackage[acronym]{glossaries}
\newcommand{\newac}{\newacronym}
\newcommand{\ac}{\gls}
\newcommand{\Ac}{\Gls}
\newcommand{\acpl}{\glspl}
\newcommand{\Acpl}{\Glspl}
\newac{speb}{SPEB}{square position error bound}
\newac[plural=EFIMs,firstplural=Fisher information matrices (EFIMs)]{efim}{EFIM}{Fisher information matrix}
\newac{ne}{NE}{Nash equilibrium}
\newac{mse}{MSE}{mean squared error}
\newac{toa}{TOA}{time-of-arrival}
\newac{snr}{SNR}{signal-to-noise ratio}
\newac{lan}{LAN}{local area network}
\newac{psd}{PSD}{positive semidefinite}
\newac{pd}{PD}{positive definite}
\newac{wrt}{w.r.t.}{with respect to}
\newac{lhs}{L.H.S.}{left hand side}
\newac{wp1}{w.p.1}{with probability 1}
\newac{kkt}{KKT}{Karush-Kuhn-Tucker}
\newac{wlog}{w.l.o.g.}{without loss of generality}
\newac{mle}{MLE}{maximum likelihood estimation}
\newac{gps}{GPS}{global positioning system}
\newac{rssi}{RSSI}{received signal strength indicator}
\newac{mimo}{MIMO}{multiple-input multiple-output}
\newac{csi}{CSI}{channel state information}
\newac{fdd}{FDD}{frequency division duplexing}
\newac{ms}{MS}{mobile station}
\newac{bs}{BS}{base station}
\newac{d2d}{D2D}{device-to-device}
\newac{slnr}{SLNR}{signal-to-interference-leakage-and-noise-ratio}
\newac{ula}{ULA}{uniform linear antenna array}
\newac{pas}{PAS}{power angular spectrum}
\newac{mmse}{MMSE}{minimum mean square error}
\newac{zf}{ZF}{zero-forcing}
\newac{rzf}{RZF}{regularized zero-forcing}
\newac{as}{AS}{angular spread}
\newac{aod}{AOD}{angle of departure}
\newac{iid}{i.i.d.}{independent and identically distributed} 
\newac{sinr}{SINR}{signal-to-interference-and-noise ratio}
\newac{tdd}{TDD}{time-division duplex}
\newac{rvq}{RVQ}{random vector quantization}
\newac{rhs}{R.H.S.}{right hand side}
\newac{mrc}{MRC}{maximum ratio combining}
\newac{cdf}{CDF}{cumulative distribution function}
\newac{a.s.}{a.s.}{almost surely}
\newac{los}{LOS}{line-of-sight}
\newac{jsdm}{JSDM}{joint spatial division and multiplexing}
\newac{map}{MAP}{maximum a posteriori}
\newac{klt}{KLT}{Karhunen-Lo\`eve Transform}
\newac{lbe}{LBE}{link bargaining equilibrium}
\newac{se}{SE}{Stackelberg equilibrium}
\newac{uav}{UAV}{unmanned aerial vehicle}
\newac{nlos}{NLOS}{non-line-of-sight}
\newac{pdf}{PDF}{probability density function}
\newac{em}{EM}{expectation-maximization}
\newac{knn}{KNN}{$k$-nearest neighbor}
\newac{svd}{SVD}{singular value decomposition}
\newac{nmf}{NMF}{non-negative matrix factorization}
\newac{umf}{UMF}{unimodality-constrained matrix factorization}
\newac{rmse}{RMSE}{rooted mean squared error}
\newac{olos}{OLOS}{obstructed line-of-sight}
\newac{mmw}{mmW}{millimeter wave}
\newac{ber}{BER}{bit error rate}
\newac{rss}{RSS}{received signal strength}
\newac{lp}{LP}{linear program}
\newac{ufw}{U-FW}{unimodal Frank-Wolfe}
\newac{utf}{UTF}{unimodality-constrained tensor factorization}
\newac{fw}{FW}{Frank-Wolfe}
\newac{iot}{IoT}{Internet-of-Things}
\newac{mae}{MAE}{mean absolute error}
\newac{crb}{CRB}{Cram\'er-Rao bound}
\newac{aoa}{AoA}{angle of arrival}
\newac{wcl}{WCL}{weighted centroid localization}
\newac[plural=GNSS,firstplural=global navigation satellite systems (GNSS)]{gnss}{GNSS}{global navigation satellite system}
\newac{gsm}{GSM}{global system for mobile communications}
\newac{imu}{IMU}{inertial measurement unit}
\newac{rbf}{RBF}{radial basis function}
\newac{msf}{MSF}{multi-sensor fusion}
\newac{lidar}{LiDAR}{light detection and ranging}
\newac{glrt}{GLRT}{generalized likelihood ratio test}
\newac{sdr}{SDR}{software-defined radio}
\newac{ap}{AP}{access point}
\newac{lte}{LTE}{Long-Term Evolution}
\newac{raim}{RAIM}{receiver autonomous integrity monitoring}
\newac{pvt}{PVT}{position, velocity and time}
\newac{npl}{NPL}{Neyman-Pearson lemma}
\newac{ekf}{EKF}{extended Kalman filter}
\newac{tdoa}{TDOA}{time difference of arrival}
\newac{gmm}{GMM}{Gaussian mixture model}
\newac{svm}{SVM}{support vector machine}
\newac{mlp}{MLP}{multilayer perceptron}
\newac{cnn}{CNN}{convolutional neural network}
\newac{lstm}{LSTM}{long short-term memory}
\newac{rnn}{RNN}{recurrent neural network}
\newac{ann}{ANN}{artificial neural network}
\newac{sop}{SOP}{signals of opportunity}
\newac{cn0}{C/N0}{carrier to noise density ratio}
\newac{agc}{AGC}{automatic gain control}
\newac{roc}{ROC}{receiver-operating characteristic}
\newac{auc}{AUC}{area under the curve}

\makeatother

\providecommand{\corollaryname}{Corollary}
\providecommand{\lemmaname}{Lemma}
\providecommand{\propositionname}{Proposition}
\providecommand{\theoremname}{Theorem}

\definecolor{mycolor1}{rgb}{0.494117647058824,0.184313725490196,0.556862745098039}
\definecolor{mycolor2}{rgb}{0.466666666666667,0.674509803921569,0.188235294117647}
\definecolor{mycolor3}{rgb}{0.301960784313725,0.745098039215686,0.933333333333333}
\definecolor{mycolor4}{rgb}{0.929411764705882,0.694117647058824,0.125490196078431}
\definecolor{mycolor5}{rgb}{0.635294117647059,0.078431372549020,0.184313725490196}
\definecolor{mycolor6}{rgb}{0.8500,0.3250,0.0980}

\begin{document}
\title{Self-supervised federated GNSS spoofing detection with opportunistic data}

\maketitle              
\begin{abstract}
\Acpl{gnss} are vulnerable to spoofing attacks, with adversarial signals manipulating the location or time information of receivers, potentially causing severe disruptions. The task of discerning the spoofing signals from benign ones is naturally relevant for machine learning, thus recent interest in applying it for detection. While deep learning-based methods are promising, they require extensive labeled datasets, consume significant computational resources, and raise privacy concerns due to the sensitive nature of position data. This is why this paper proposes a self-supervised federated learning framework for \ac{gnss} spoofing detection. It consists of a cloud server and local mobile platforms. Each mobile platform employs a self-supervised anomaly detector using \ac{lstm} networks. Labels for training are generated locally through a spoofing-deviation prediction algorithm, ensuring privacy. Local models are trained independently, and only their parameters are uploaded to the cloud server, which aggregates them into a global model using FedAvg. The updated global model is then distributed back to the mobile platforms and trained iteratively. The evaluation shows that our self-supervised federated learning framework outperforms position-based and deep learning-based methods in detecting spoofing attacks while preserving data privacy. 
\end{abstract}


\glsresetall

\section{Introduction}
\Acpl{gnss} provided location and time information is integrated into many aspects of everyday life, with applications ranging from autonomous vehicles to mobile map navigation. However, \ac{gnss} is vulnerable to spoofing attacks, with false satellite signals tampering with the location or time information of the \ac{gnss} receivers. This can result in misleading navigation \cite{Goo:J22}, making an autonomous vehicle crash \cite{SheWonCheChe:C20}, or disrupting timing, e.g., in power systems \cite{BinZiwYonLia:J20}. Furthermore, attack sophistication can range from ones mounted with single relatively simple devices to multiple sophisticated spoofers \cite{LiuCheYanShu:C21}. 

A multiplicity of methods has been proposed to defend \ac{gnss} receivers, including \ac{raim}, signal processing, and statistical testing \cite{KhaRosLanCha:C14,MaaKas:J21,RotCheLoWal:J21,LiuPap:C23}. Such schemes can be effective in detecting attacks utilizing Doppler shift, \ac{snr}, \ac{sop}, and \ac{imu}, as these opportunistic data and signal properties inherently assist the detection process. These detectors are typically implemented at individual \ac{gnss} receivers or their encompassing platforms, with little consideration of privacy preservation or the utilization of distributed data across multiple devices. Operating in diverse volatile settings, with complex radio propagation environments, while facing adversaries, can be challenging, with varying attack detection performance. This calls for data-driven methods to improve accuracy and reliability. 

A promising approach to enhance detection is to collect data from mobile platforms with \ac{gnss} and leverage machine learning to train detection models. Deep learning-based methods \cite{MorDelLoh:J19,BorLiWuClo:J24} can use signal properties and opportunistic data as input features, e.g., power, phase, and \ac{snr}. These methods need a large number of labeled datasets, which are costly and labor-intensive in real-world environments. Moreover, they require significant processing power on the server side. Beyond these challenges, the collection and upload of data to the server for training raises privacy concerns. Position data is highly sensitive, which limits the sharing of data among \ac{gnss} receivers and restricts the development of collective, robust detection methods; thus, federated learning \cite{WuCalImbClo:C23,DenLuoYao:C24} is proposed to enhance privacy and detect \ac{gnss} attacks. 

While recent federated learning approaches \cite{WuCalImbClo:C23,DenLuoYao:C24,NguMarMieFer:C19,FerDmiRieMie:C22} can be used for attack detection, none of the existing works have explored self-supervised federated learning for \ac{gnss} attack detection. Since labeling datasets is labor-intensive, self-supervision is important for enabling real-time and adaptive online training, offering greater practicality for real-world applications. Therefore, the objective of this paper is to investigate a self-supervised federated learning framework based on our previous \ac{gnss} attack detection scheme \cite{LiuPap:C23}, which locally generates spoofing likelihood as labels from its detector by using \ac{sop} and \ac{imu}. A key challenge is how to use these labels with \ac{gnss}, opportunistic data, and signals’ properties to train local models across mobile platforms. Another challenge is how to federate and aggregate the local models from different mobile platforms into a stronger global model for detection. In addition, due to privacy concerns, sharing datasets containing positional, motion, and network-related information with other mobile platforms or server is prohibited by design. 

Our proposed framework contains two main parts. The server part is a cloud server responsible for tasks such as housing the global detector model and model aggregation. The edge clients part involves numerous local mobile platforms, each equipped with a self-supervised anomaly detector model. These models utilize \ac{lstm} networks, which are also well-suited for capturing temporal dependencies in sequential data. Each platform processes a local dataset, which includes time-series information collected from \ac{gnss} receiver (\ac{gnss} position, \ac{agc}, \ac{cn0}, Doppler shift, etc.), network infrastructures (network position), and onboard sensors (acceleration, attitude, etc.). The self-supervised model is trained using the spoofing-deviation labels generated from the method in \cite{LiuPap:C23}. The local mobile platforms do not upload their local datasets to the cloud server. Instead, they transmit only the parameters of their respective local models. Upon receiving these parameters, the cloud server performs model aggregation by combining them into a single global model using the FedAvg algorithm. The aggregated global model is distributed back to the mobile platforms and then trained again.

In the experimental evaluation, we test \ac{gnss} spoofing data that we collected from Jammertest 2024 \cite{Jam:J24} using multiple vehicle-mounted smartphones. The dataset includes both \ac{iid} and non-\ac{iid} \ac{gnss} positions and opportunistic data across different mobile devices---the former refers to training and testing data originating from the same trace and device, while the latter involves data from different traces or devices. We found that the proposed self-supervised federated learning and its corresponding centralized learning have performance gains over the baseline. Meanwhile, the proposed federated method preserves position privacy for mobile platforms. 

The novelty and contributions of this paper are: 
\begin{itemize}
    \item Utilization of self-supervised federated learning with opportunistic data for \ac{gnss} spoofing detection. 
    \item Elimination of the requirement for labeled or annotated datasets in deep learning-based \ac{gnss} spoofing detection. 
    \item Performance gain over position-based and deep learning-based methods on real \ac{gnss} spoofing detection, in terms of accuracy. 
\end{itemize}
It is also noteworthy that the above are achieved without user privacy deterioration compared with the deep learning method with centralized training. 

The remainder of the paper is organized as follows: Sec.~\ref{sec:relwor} provides background and reviews related work for \ac{gnss} attacks and learning-based countermeasures. Sec.~\ref{sec:sysmod} presents our system model and adversary. Sec.~\ref{sec:prosch} details the problem and the proposed scheme. Sec.~\ref{sec:numres} discusses evaluation and comparison with baseline methods. Finally, Sec.~\ref{sec:conclu} concludes the paper.

\section{Related Work and Background}
\label{sec:relwor}
This section provides an overview of \ac{gnss} attacks and reviews related work of attack detection using machine learning techniques, including deep learning. Subsequently, we summarize recent advancements in federated learning with a focus on security and privacy. 
\subsection{GNSS Jamming and Spoofing Attacks}
\Ac{gnss} jamming disrupts receivers by transmitting high-power radio frequency signals within or near \ac{gnss} frequency bands, effectively overpowering the legitimate satellite signals. \Ac{gnss} spoofing maliciously manipulates user position and time, especially because civilian \ac{gnss} usually lacks authentication, and the protocol, encoding, and modulation are publicly available. Even if \ac{gnss} signals were authenticated, the attacker can launch a relaying or replaying attack on \ac{gnss} \cite{LenSpaPap:C21}. Prior to spoofing, \ac{gnss} jamming is often used to force the \ac{gnss} receiver out of the satellite signal lock \cite{KasKhaAbdLee:J22}. The simplest way of generating a spoofing signal is meaconing, which is the retransmission of legitimate \ac{gnss} signals with a time delay \cite{LenSpaPap:C21}. However, if the receiver has an accurate timer and already knows its recent location, the delayed time introduced by the meaconer will result in a sudden time shift, which may be detected. A variation of meaconing, selective delay, can rebroadcast individual satellite signals \cite{PapJov:C08}. This can modify the position solution only without changing the time. Other sophisticated spoofing attacks can overcome more technical limitations, such as portable spoofers, which are attached to the victim \cite{SchRadCamFoo:J16}. Additionally, \cite{NarRanNou:C19} focuses on the strategy of \ac{gps} spoofing, which combines the road contextual information of the city map and can generate a designed route for spoofing a moving receiver. 

\subsection{Artificial Intelligence for GNSS Security}
Machine learning, particularly deep learning, against \ac{gnss} attacks has shown promising potential and gained momentum, utilizing a range of models such as random forest \cite{XuYinLi:J20}, \ac{svm} \cite{MorDelLoh:J19,XuYinLi:J20}, \ac{mlp} \cite{BorLiWuClo:C20,TohMos:C20,BorLiWuClo:J24}, \ac{cnn} \cite{BorLiWuClo:C20,KarPalJay:C21,ElaUjaRuo:C22}, \ac{gmm} \cite{FenSeoCao:C22}, \ac{lstm} \cite{CalBhaBovGiu:C20,KarPalJay:C21}, and \ac{rnn} \cite{CalBhaBovGiu:C20}. 

For detecting \ac{gnss} jamming, \cite{XuYinLi:J20} focuses on analyzing three common \ac{gnss} interference signals by extracting various entropy features (e.g., power spectral entropy), creating a combined entropy dataset, and utilizing \ac{svm} and radio frequency methods to classify the signals. Alternatively, \cite{MorDelLoh:J19} treats the classification of jammers in \ac{gnss} bands as a black-and-white image classification problem. Time-frequency analysis and image mapping of jammed signals are used to categorize the received signal into six classes. This method achieves notable classification accuracies of up to 94.90\% with \ac{svm} and up to 91.36\% with \ac{cnn}. 

To detect \ac{gnss} spoofing, \cite{BorLiWuClo:C20} explores a cross-ambiguity function to train data-driven models for probabilistic classification, which focuses on each satellite individually and makes use of complex neural networks, including an \ac{mlp} and two types of \acpl{cnn}. In \cite{FenSeoCao:C22}, a \ac{gmm}-based unsupervised method detects and mitigates \ac{gnss} signal spoofing by clustering the positions generated by the benign \ac{gnss} signals and isolating spoofed pseudoranges. Furthermore, \cite{KarPalJay:C21} utilizes both \ac{cnn} and \ac{lstm} to identify a spoofer by classifying the pairwise cross-correlation of different receivers and comparing the cyclic profiles, then observes that \ac{cnn} achieves the highest accuracy. \ac{lstm} for anomaly detection leverages the predictability of the Doppler traits of the received \ac{gnss} signals: the data was collected using cost-effective \ac{sdr} receivers and processed on affordable embedded platforms (e.g., Jetson Nano) to predict Doppler shift for spoofing detection \cite{CalBhaBovGiu:C20}. 

When \ac{gnss} signals are mixed up with jamming, spoofing, and other interference signals, a robust deep learning technique combined pre-trained \ac{cnn} with transfer learning to detect and classify disruptions of \ac{gnss} signals based on time-frequency analysis in \cite{ElaUjaRuo:C22}. An \ac{ann} trained by particle swarm optimization is proposed to detect various types of interactions affecting \ac{gnss} \cite{TohMos:C20}. By using received signal power and distortion in the correlation function as feature vectors, this \ac{ann} classifies received signals into categories such as jammed, spoofed, multi-path afflicted, or interference-free. \cite{KaaMakRuoMal:C21} introduces the \ac{gnss}-Finland monitoring platform, which employs the FinnRef reference network and deep learning methods to analyze big data from \ac{gnss}-Finland to identify trends in signal quality, detect anomalies, assess continuity, and forecast crucial failures in positioning and timing. 

However, the limitations of these machine learning methods come when high-quality training data with annotation is unavailable. Additionally, when processing non-\ac{iid} datasets, the offline deep learning models face challenges in achieving generalization.

\subsection{Federated Learning for Security and Privacy}
Federated learning \cite{McmMooRamHam:C17} has emerged as a promising approach for training machine learning models while addressing privacy concerns. Instead of transferring the raw dataset to a central server for training, participating devices collaboratively train the model locally and share only the model updates (gradients or weights) with the central server. 

Considering the application of federated learning in security, since the increasing deployment of \ac{iot} devices in daily life has resulted in many vulnerable devices, yet existing intrusion detection techniques are ineffective due to the massive scale of the problem and diverse types of devices and manufacturers involved. Therefore, \cite{NguMarMieFer:C19} introduces an autonomous self-learning distributed system that utilizes device-type-specific communication profiles to detect anomalous deviations in communication without human intervention or labeled data, leveraging a federated learning approach for efficient profile aggregation, making it the first system to employ this approach for intrusion detection based on anomaly detection. FedCRI in \cite{FerDmiRieMie:C22} is a solution for sharing cyber-risk intelligence, wherein mobile cyber-risks were transformed into effective risk detection models based on contributions from different mobile service providers, and their extensive evaluation on real-world user databases representing 23.8 million users of security-critical mobile apps, enabling effective identification of risks on mobile devices.

Federated learning has also proven its success in privacy-preserving applications beyond security. To improve next-word prediction in smartphone virtual keyboards, Google deploys an \ac{rnn} language model trained using federated learning \cite{HarRaoMatRam:J18}, a distributed on-device learning framework and the effectiveness of server-based training with stochastic gradient descent is compared to client device training, which showcases the advantage of training language models on client devices without compromising user data privacy and gives users more control over their data usage. Similarly, for object detection in autonomous driving systems, \cite{JalRavBadUch:C21} proposes a federated learning-based approach that preserves data privacy while maintaining performance by training the model in a decentralized manner and analyzes the impact of this decentralized approach on object detection performance in a real-world traffic environment. In addition, the challenge of deep learning-based medicine lies in finding sufficiently large and diverse datasets, which are rare in individual institutions, leading to privacy and ownership challenges. Then, in \cite{SheEdwReiMar:J20}, through a paradigm for data-private multi-institutional collaborations, models trained among 10 institutions achieve 99\% of the quality achieved with centralized data. 

Despite its advantages, federated learning is vulnerable to adversarial attacks. Current defenses either rely on techniques such as differential privacy or analyze model weights using outlier detection methods limited to specific data distributions, so \cite{RieKraMieDmi:J22} proposes CrowdGuard, a model filtering defense that leverages client data and secure enclaves to analyze individual models without data leaks. It introduces a novel metric to analyze network hidden layer outputs, coupled with a significance-based detection algorithm, enabling effective detection of poisoned models even in non-\ac{iid} scenarios. Metric-Cascades in \cite{KraDmi:J23} use multiple detection metrics, such as Euclidean magnitude and direction, to filter poisoned model updates. The evaluation demonstrated that it successfully distinguishes backdoors from distortions, making it the first defense resilient to strong adaptive adversaries in real-world scenarios with minimal overhead. On the other hand, a malicious server can also use uploaded models to derive sensitive information. Hence, a decentralized framework is proposed in \cite{MorRamPanMon:J22} that utilizes multi-party computation primitives like secret sharing, providing strict privacy guarantees against curious aggregators or colluding data owners.

\section{System Model and Adversary}
\label{sec:sysmod}
\begin{figure}
\begin{centering}
\includegraphics[width=\columnwidth]{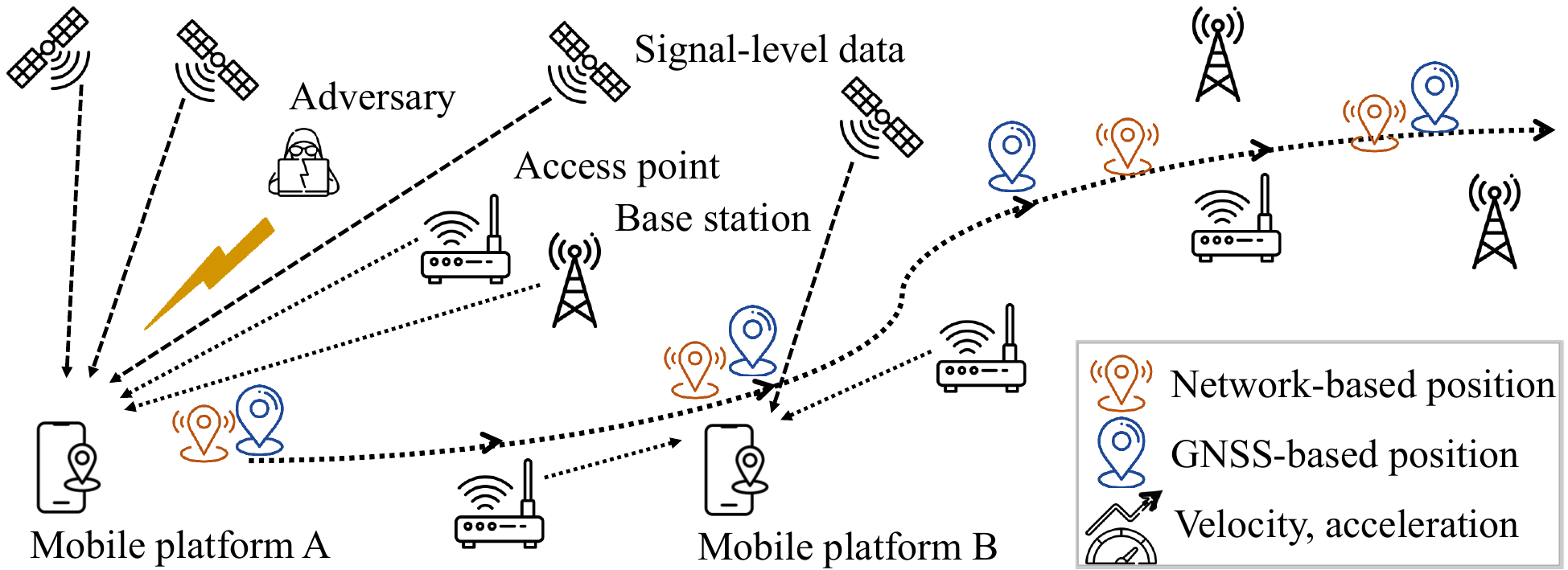}
\par\end{centering}
\caption{System and adversary model illustration.}
\label{fig:sysmod}
\end{figure}

As shown in Fig.~\ref{fig:sysmod}, we consider multiple mobile \ac{gnss} platforms (e.g., smartphone, car, and drone) equipped with common modules (e.g., Wi-Fi, Bluetooth, cellular, \ac{imu}, and speed sensors), and streaming signal-level properties from satellites (\ac{agc}, antenna \ac{cn0}, baseband \ac{cn0}, and Doppler shift). The platforms are connected to a cloud server, which is curious but honest, so sharing datasets that contain location, motion, and network information with the cloud server or other mobile platforms is not acceptable. 

Provided that a \ac{gnss} position attack is available, when a mobile platform moves and navigates within the attack area, the \ac{gnss}-provided location will deviate from the actual location. As a result, network-provided positions, motion information, and signal-level properties will be inconsistent with \ac{gnss} position information and its benign behavior. However, the training data (encompassing the \ac{gnss} receiver, network infrastructures, and onboard sensors) is not manually labeled or annotated to indicate ``benign'' or ``under attack''. Mobile platforms may differ, i.e., they may be navigating in different areas and using different hardware. This results in a so-called non-\ac{iid} data situation. Similarly, if the hardware were the same and the navigation traces were similar, the data could be considered \ac{iid}. 

The external adversary uses spoofed or replayed/relayed \ac{gps} and/or Galileo signals to force \ac{gnss} receivers to incorrectly compute their positions (and/or time) \cite{ZhaLarPap:J22,LenSpaPap:C21,LenSpaPap:C22,SpaPap:J24}. We assume that the adversary knows the victim location and can use state-of-the-art attack techniques. We assume the attacker only operates within the \ac{gnss} domain and does not attack other network infrastructures, including cellular and Wi-Fi communications. Additionally, we assume the attacker does not have physical control over the server and mobile platforms, and thus cannot manipulate the process of deriving detection models and model parameters. 

\textbf{Notations.} We denote the position of the $m$th mobile platform at time $t$ is located at the coordinates $\mathbf{p}_{\text{true}}(m,t) \in \mathbb{R}^2$, where $m=1,2,...,M$; $M$ is the total number of mobile platform. The said platform has a \ac{gnss} provided position $\mathbf{p}_{\text{gnss}}(m,t)$, a network-provided position $\mathbf{p}_{\text{net}}(m,t)$, as well as speed $\mathbf{v}(m,t)$, acceleration $\mathbf{a}(m,t)$, attitude $\boldsymbol{\omega}(m,t)$ from onboard sensors, and \ac{gnss} signal properties $\mathbf{s}(m,t)$, which encompasses mean, median, minimum, and maximum values of \ac{agc}, antenna \ac{cn0}, baseband \ac{cn0}, and Doppler shift. 

\begin{figure*}
\begin{centering}
\includegraphics[trim={0 0 0 2.5cm},clip,width=1.55\columnwidth]{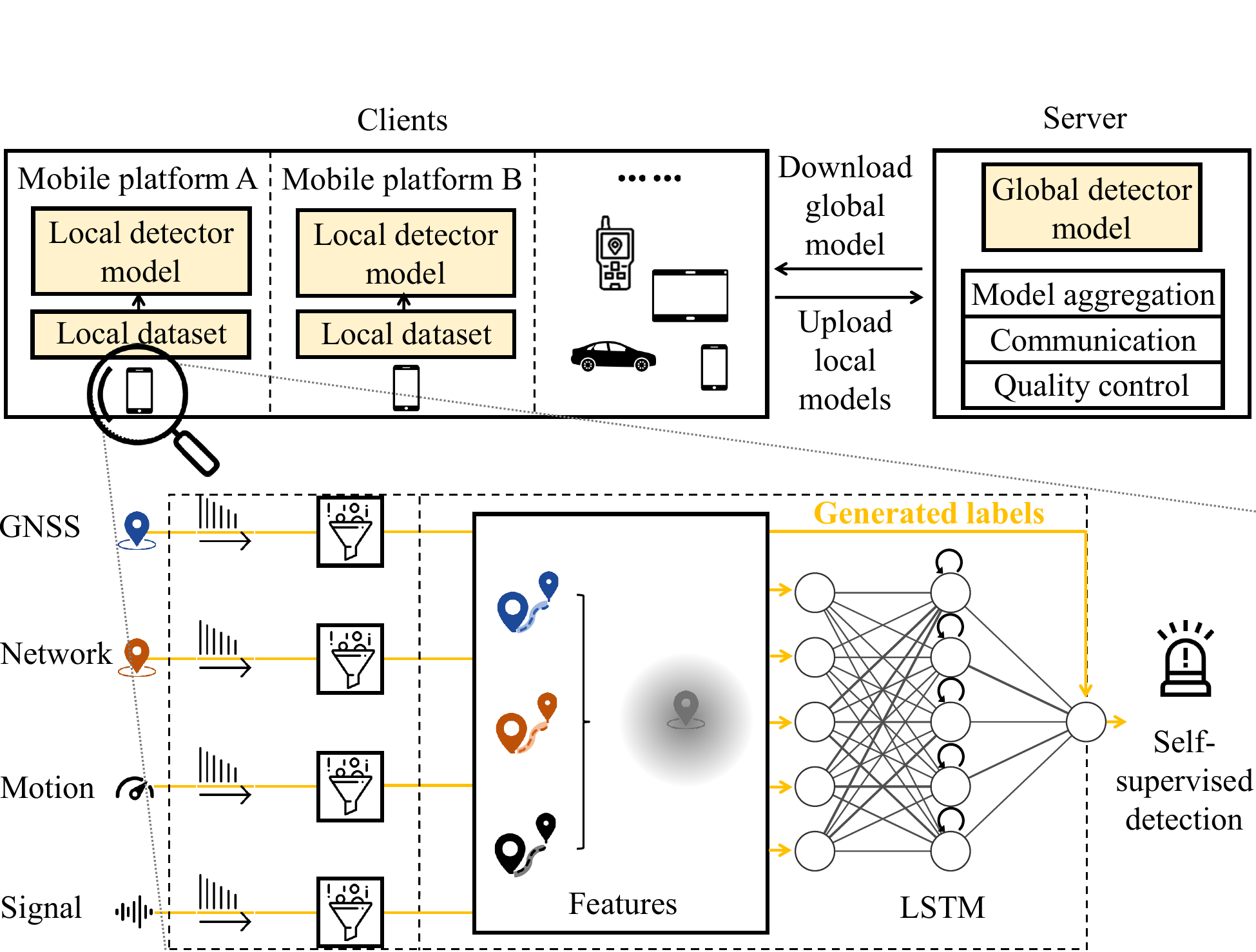}
\par\end{centering}
\caption{Overview of self-supervised federated \ac{gnss} spoofing detection.}
\label{fig:scheme}
\end{figure*}

\section{Proposed Scheme}
\label{sec:prosch}
The proposed scheme uses a federated self-supervised learning-based detection framework, as depicted in Fig.~\ref{fig:scheme}. This framework allows for effective collaboration and online knowledge sharing among mobile platforms while safeguarding the privacy of user data. 

The mobile platforms collect $\mathbf{p}_{\text{gnss}}(m,t)$, $\mathbf{p}_{\text{net}}(m,t)$, $\mathbf{v}(m,t)$, $\mathbf{a}(m,t)$, $\boldsymbol{\omega}(m,t)$, and $\mathbf{s}(m,t)$ to construct local datasets (Sec.~\ref{subsec:feaeng}). Each platform trains an \ac{lstm} regression model in Sec.~\ref{subsec:modstr} using its local dataset as input with the generated labels from Sec.~\ref{subsec:labgen}, and then uploads the model parameters to the cloud server. The cloud server adopts the FedAvg algorithm to aggregate model parameters trained based on the distributed datasets (Sec.~\ref{subsec:modagg}) to improve both the accuracy of spoofing detection and data privacy. 

\subsection{Feature Engineering}
\label{subsec:feaeng}
On each mobile platform, we construct two kinds of features from the local dataset: position-based and signal-based features. 

The position-based features stem from our previous scheme \cite{LiuPap:C23}, which provides a secure fused position $\mu \in \mathbb{R}^2$ with uncertainty $\sigma \in \mathbb{R}^2$ using $\mathbf{p}_{\text{gnss}}(m,t)$, $\mathbf{p}_{\text{net}}(m,t)$, $\mathbf{v}(m,t)$, $\mathbf{a}(m,t)$, and $\boldsymbol{\omega}(m,t)$. Four elements are encompassed: the estimated position residual and uncertainty of latitude and longitude, calculated by $\{\mu(m,t)-\mathbf{p}_{\text{gnss}}(m,t), \sigma(m,t)\} \in \mathbb{R}^4$. The signal-based features are four statistics of satellite signals. We calculated mean, median, minimum, and maximum values of the following physical properties for \ac{gps} L1 and Galileo E1 (can be extended to other constellations):
\begin{itemize}
    \item \Ac{agc}: Regulates signal amplitude by automatically adjusting the receiver’s gain to compensate for variations in signal power, which can also indicate interference.
    \item Antenna \ac{cn0}: Represents the ratio of the signal power to the noise power density at the antenna, influenced by atmospheric conditions, satellite elevation, and interference.
    \item Baseband \ac{cn0}: Measures the signal quality after down-conversion and filtering, similar to antenna \ac{cn0}.
    \item Doppler shift: Captures frequency changes caused by satellite-receiver relative motion, essential for satellite fingerprint construction and velocity estimation.
\end{itemize}
In a benign environment, \ac{agc}, \ac{cn0}, and Doppler shift of different satellites should be different for each satellite. However, spoofing often causes artificially similar values and typically higher signal power. These signal-based features have $4 \times 4 \times 2$ elements, corresponding to four types of statistics, four properties, and two constellations. Hence, the extracted feature set includes 32 signal-based elements and four elements representing the estimated position residual and uncertainty in latitude and longitude, in a total of 36 elements. 

Moreover, features are normalized and cleaned within the local dataset. For the position-based features, the estimated position residual and uncertainty may contain extreme values. To mitigate extremes, values exceeding the 95th percentile are capped at the 95th percentile threshold. In the signal-based feature, some signal properties contain invalid values (e.g., missing or faulty measurements). These invalid values are replaced with the minimum value within their respective valid range. Finally, a min-max scaling strategy is applied separately to each feature type. Each feature is rescaled to the range $[0,1]$ to ensure stable training. 

\subsection{Label Generation}
\label{subsec:labgen}
Self-supervised labels are generated based on the method proposed in \cite{LiuPap:C23}. The generation uses $\mathbf{p}_{\text{gnss}}(m,t)$, $\mathbf{p}_{\text{net}}(m,t)$, $\mathbf{v}(m,t)$, $\mathbf{a}(m,t)$, and $\boldsymbol{\omega}(m,t)$ to compute a scalar value of the estimated \ac{gnss} position deviation, normalized to the range $[0,1]$, for each $(m,t)$ in a fully automated manner. 

We first compute the Euclidean norm of the estimated position residual, $\mu(m,t)-\mathbf{p}_{\text{gnss}}(m,t)$ (the difference between the secure fused position in \cite{LiuPap:C23} and \ac{gnss} position), to obtain a scalar representation of the estimated \ac{gnss} deviation. To remove extreme values, deviations exceeding the 95th percentile are capped at this threshold. Finally, we apply min-max scaling to normalize the values. 

These normalized values serve as self-supervised labels, representing the estimated \ac{gnss} position deviation at each $(m,t)$. As they are generated entirely locally without any external manual annotation, this approach enables self-supervised learning. Note that the algorithm used for the generation of these labels is not a contribution of this work; instead, we use these labels to build our self-supervised federated learning for \ac{gnss} spoofing detection. 

\subsection{Model Structure}
\label{subsec:modstr}
We implement the detection model based on an \ac{lstm} neural network on each mobile platform, as \ac{lstm} is well-suited for processing time series data. The network input is the features from Sec.~\ref{subsec:feaeng}, and the output is a scale value in $[0,1]$. The objective is to capture the temporal dependencies of each signal feature and identify anomalies. 

\textbf{Layer structure.} The model contains two \ac{lstm} layers and a fully connected output layer. The first \ac{lstm} layer contains 100 units, which takes time series feature input and outputs a sequence of hidden states at each moment. The second \ac{lstm} layer also contains 100 units but processes the sequence of hidden states from the first layer and outputs the hidden state at the last time step only. 

\textbf{Activation function.} In the last fully connected layer, we use the Sigmoid activation function to output a probability value to indicate whether the current position is a result of \ac{gnss} spoofing. 

\textbf{Loss function.} The loss function of the model uses the \ac{mse} to reduce the prediction error of deviation. 

\textbf{Batch size and learning rate.} The batch size of the model is set to 72; the learning rate is adjusted proportionally to adapt to the different data volumes of each device. 

\textbf{Early stopping strategy.} The EarlyStopping strategy is used in training. When the verification loss does not improve after 20 epochs, the training is stopped to avoid overfitting. 

\subsection{Model Aggregation}
\label{subsec:modagg}
Mobile platforms do not upload local datasets but upload trained local model parameters. The cloud server aggregates these local model parameters and uses the FedAvg algorithm to average them to form a global model. The server sends global model parameters to all devices after each iteration to achieve continuous learning. It performs multiple iterations of aggregation on the local dataset of different devices to improve the generalization ability of the model.

\subsection{Quality Control}
After the global model undergoes sufficient rounds of federated training, its detection accuracy should stabilize. At this point, the system initiates a quality control process to assess parameter updates from each local model. In each learning iteration, upon receiving a local model update, the cloud server distributes this update to other participating mobile platforms for independent evaluation on their respective local datasets. Then, the mobile platforms upload the predicted labels, and the server computes a performance metric---specifically, the \ac{auc} score---based on these uploaded labels and the predictions from the previous global model. The FedAvg algorithm will only accept this parameter update (from a local model) if the metric is better than a criterion threshold. Given that we only consider all participants who contribute model parameters that are benign, the quality control here aims to filter out low-quality data samples. 

\begin{figure}
\begin{centering}
\includegraphics[width=\columnwidth]{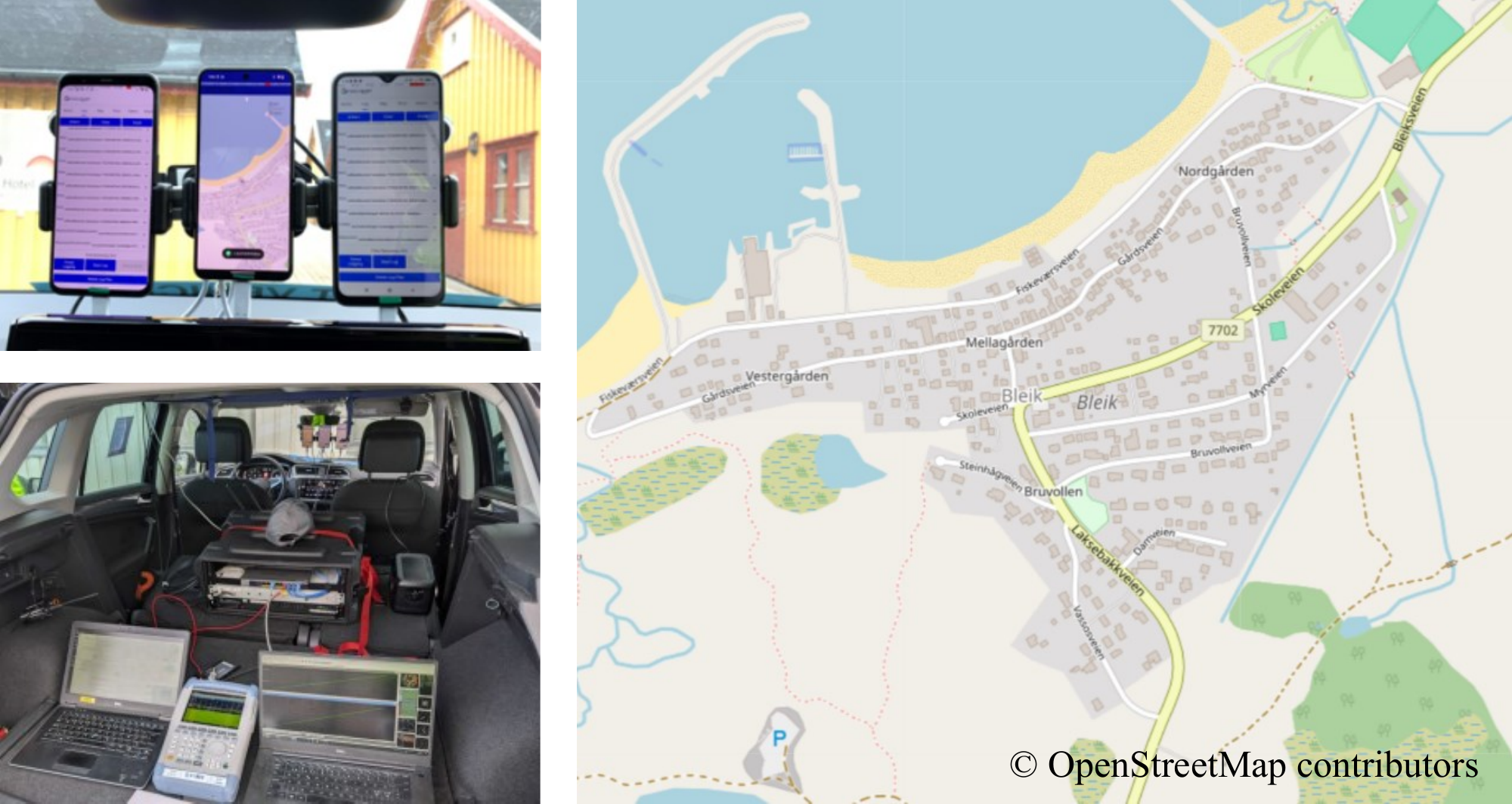}
\par\end{centering}
\caption{Jammertest main test area (right) and mounted smartphones in a vehicle (left).}
\label{fig:dataset}
\end{figure}

\section{Experimental Evaluation}
\label{sec:numres}
As shown in Fig.~\ref{fig:dataset}, our \ac{gnss} spoofing dataset is sourced from Jammertest 2024 \cite{Jam:J24}, and includes data collection from six Android smartphones. Phone 1 and Phone 4 are Google Pixel 8. Phone 2 and Phone 6 are Google Pixel 4 XL. Phone 3 is Xiaomi Redmi 9, and Phone 5 is Samsung Galaxy S9. All smartphones support multiple constellations, while only Google Pixel 4 XL and Pixel 8 support double frequencies. The dataset consists of 85 drive-testing traces in total that were recorded throughout the day, from morning to evening, around Bleik town. The smartphones record timestamps, \ac{gnss} positions, network positions, \ac{imu} data, and \ac{gnss} signal properties (\ac{agc}, antenna \ac{cn0}, baseband \ac{cn0}, and Doppler shift) via Android APIs. Ground truth positions are obtained from two u-blox ZED-F9P receivers using benign constellations with the help of a nearby reference station. Labels for self-supervised learning are generated locally through the opportunistic data fusion algorithm \cite{LiuPap:C23}, as Sec.~\ref{subsec:labgen}. 

The deep learning framework is Keras with a TensorFlow backend, and we choose \ac{lstm} module to process time series data, so it can capture temporal dependencies in features. Regarding the federated learning part, we implement the FedAvg algorithm on the cloud server to average all local model parameters. Additionally, we do not simulate processing overhead, transmission delays, or packet loss in wireless communications between the server and clients. 

We have different train/test data-splitting methods. One is to split the data based on different smartphones without considering different traces. Another is to randomly choose 10\% of traces for testing, while the rest of the traces are for training. In addition, as our detection is self-supervised, we use the entire dataset for training and testing for the comparison of centralized or federated detection. 

\subsection{Evaluation Metrics}
To evaluate the performance of the proposed scheme, we use true positive rate ($R_\mathrm{TP}$) versus false positive rate ($R_\mathrm{FP}$) and plot \ac{roc} curves. Additionally, we calculate \ac{auc} values, which are the area under the \ac{roc} curves. To systematically evaluate different scenarios, our experiments analyze the metrics across same-device training and testing, same-model training and testing, and cross-model generalization using same or different traces.

\begin{figure}
\begin{centering}
\includegraphics[width=\columnwidth]{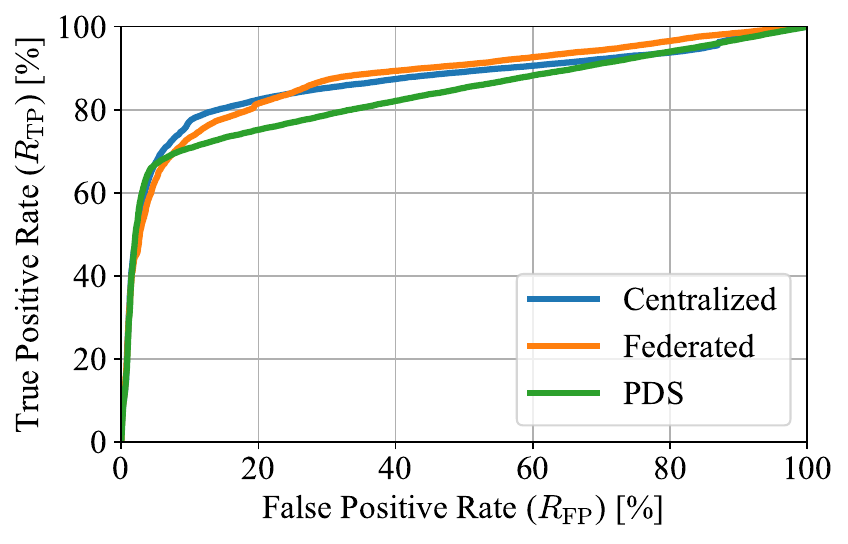}
\par\end{centering}
\caption{\ac{roc} curves of the centralized and federated self-supervised detection, and the position-based detection.}
\label{fig:trupos_clfl}
\end{figure}

\begin{figure}
\begin{centering}
\includegraphics[width=\columnwidth]{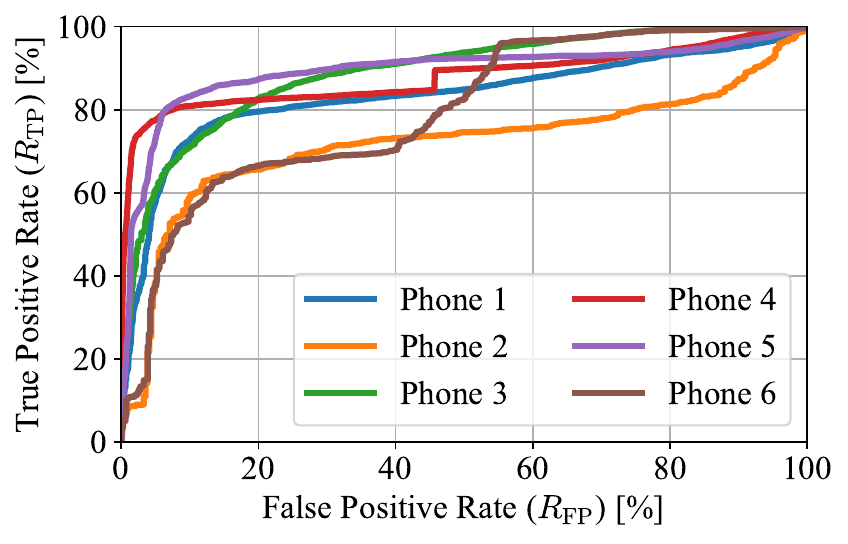}
\par\end{centering}
\caption{\ac{roc} curves of the proposed self-supervised detection for each smartphone based on their local \ac{iid} data.}
\label{fig:trupos_2.1}
\end{figure}

\subsection{Evaluation Results}
\subsubsection{Comparison Between Centralized and Federated}
We compare the detection accuracy of the proposed self-supervised detection in a centralized or federated manner with the position-based detection scheme (PDS) \cite{LiuPap:C23}. \ac{roc} curves for the methods of position fusion, centralized training, and the proposed federated learning are shown in Fig.~\ref{fig:trupos_clfl}. We find \ac{roc} curve and training loss value of the proposed become stable after 600--1000 epochs. The centralized training curve has an \ac{auc} value of 86.6\%. The federated learning curve's \ac{auc} is 87.4\%, while the curve of PDS is 83.5\%.

This comparison shows that the self-supervised learning-based approach outperforms the baseline method, and validates that federated learning can achieve competitive performance compared to centralized training. Furthermore, the proposed scheme has at most 10\% true positive rate gain over PDS \cite{LiuPap:C23} in Fig.~\ref{fig:trupos_clfl}, and preserves location data privacy. 

In our scheme, centralized training achieves a higher accuracy than federated one, when $R_\mathrm{FP}<10\%$. This phenomenon is common in federated learning due to data samples being sequential and shuffled across batches in centralized training. In contrast, federated learning involves parallel training across clients, where each client trains on relatively unrefined models and locally available data. As a result, the model struggles to generalize effectively by capturing shared patterns across the entire dataset. 

\begin{figure}
\begin{centering}
\includegraphics[width=\columnwidth]{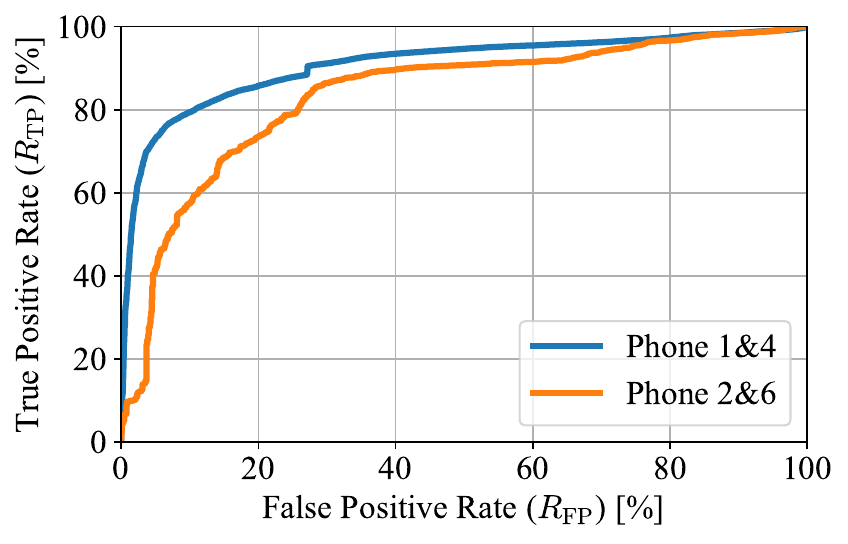}
\par\end{centering}
\caption{\ac{roc} curves of the proposed self-supervised federated detection for each smartphone model. Training and testing data are collected from the same smartphone model and driving traces.}
\label{fig:trupos_2.2}
\end{figure}

\begin{figure}
\begin{centering}
\includegraphics[width=\columnwidth]{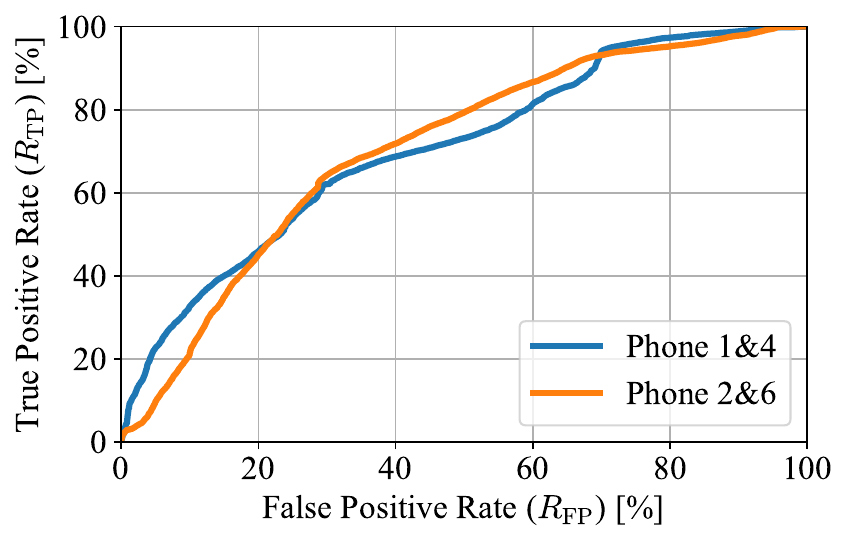}
\par\end{centering}
\caption{\ac{roc} curves of the proposed self-supervised federated detection for different smartphone models. Train/test data-splitting is based on smartphone model: Training data is collected from one smartphone model and testing data from another model, both obtained along the same driving traces.}
\label{fig:trupos_2.3}
\end{figure}

\subsubsection{Generalization Between Different Devices}
This evaluation aims to assess the generalization performance of the proposed method across different devices, both within individual devices and among devices of the same smartphone model. 

First, we compare the detection accuracy of the proposed self-supervised detection within every individual phone, termed one device training and same device testing. The result is shown in Fig.~\ref{fig:trupos_2.1}. The phones will not upload data; instead, they conduct training and testing of detectors separately and locally. Phone 1 and Phone 4 are the exact smartphone model, but the data of collected driving traces are different (non-\ac{iid}). Phone 2 and Phone 6 are also the same smartphone model, while Phone 3 and Phone 5 refer to different smartphone models. We can observe that the true positive rate of our algorithm for the Google Pixel 4 XL phones is not as good as other phones. These varying levels of performance across different phone models are potentially due to hardware differences or non-\ac{iid} data distributions. 

Then, we conduct self-supervised federated detection within devices of the same smartphone model name, termed one model training and same model testing. The result is shown in Fig.~\ref{fig:trupos_2.2}. Compared with the results of individual device training and testing in Fig.~\ref{fig:trupos_2.1}, federated detection within a smartphone model has higher accuracy, highlighting the benefits of collaborative training. 

Similarly, we attempt to train the detection using the devices of one smartphone model and test the detection using other smartphone models, termed one model training and different model testing. The result is shown in Fig.~\ref{fig:trupos_2.3}. The curve of Phone 1\&4 uses Google Pixel 8 phones for training and all other phones for testing; the curve of Phone 2\&6 uses Google Pixel 4 XL phones for training and all other phones for testing. We observe that the detector cannot perform well on a given hardware model if it has not been trained on data from the same model. 

\begin{figure}
\begin{centering}
\includegraphics[width=\columnwidth]{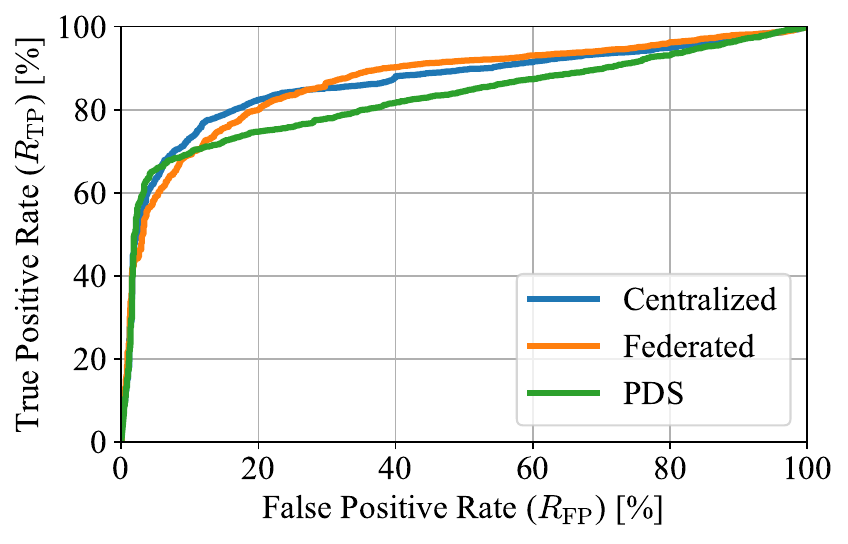}
\par\end{centering}
\caption{\ac{roc} curves of the centralized and federated self-supervised detection, and the position-based detection. Train/test data-splitting is based on driving trace: 10\% of traces are randomly chosen for testing, while the rest of the traces are for training.}
\label{fig:trupos_3.0clfl}
\end{figure}

\begin{figure}
\begin{centering}
\includegraphics[width=\columnwidth]{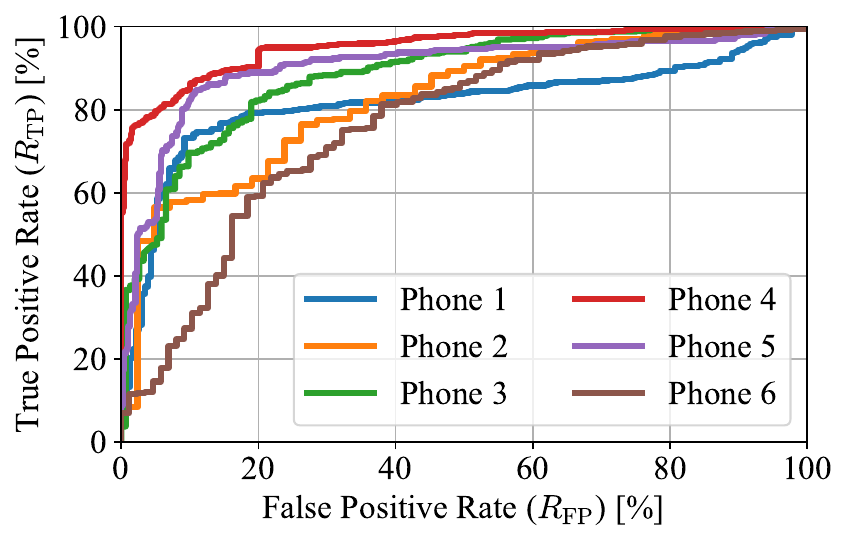}
\par\end{centering}
\caption{\ac{roc} curves of the proposed self-supervised detection for each smartphone. Train/test data-splitting is based on driving trace. Specifically, for each curve, training and testing data are from the same smartphone but with different traces.}
\label{fig:trupos_3.1}
\end{figure}

\subsubsection{Generalization Between Different Traces}
Different from the previous comparisons that use the same trace for both training and testing, this experiment divides the traces into training traces and testing traces. 

In Fig.~\ref{fig:trupos_3.0clfl}, we compare the performance of the proposed self-supervised detection in a centralized or federated manner. The centralized training curve has an \ac{auc} value of 86.5\%. The federated learning curve is 86.7\%, while the PDS curve is 83.4\%. These \ac{auc} values and their corresponding curves closely match those depicted in Fig.~\ref{fig:trupos_clfl}, i.e., 0.7\% difference at most. This indicates that the proposed method is good at generalizing between different traces. 

Likewise, we compare the detection accuracy of the proposed self-supervised detection within every individual phone, but with different traces for training and testing. The results in Fig.~\ref{fig:trupos_3.1} show similar true positive rate curves and relative performance gains to those in Fig.~\ref{fig:trupos_2.1}. 

\begin{figure}
\begin{centering}
\includegraphics[width=\columnwidth]{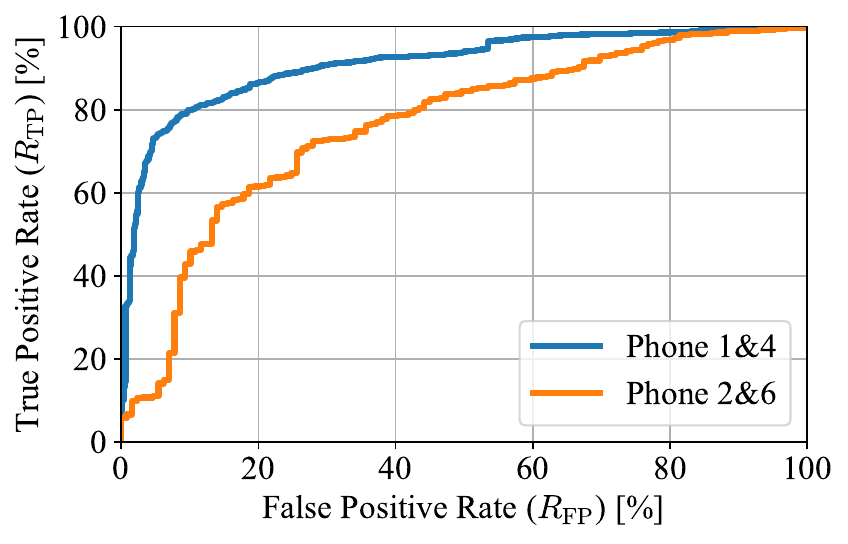}
\par\end{centering}
\caption{\ac{roc} curves of the proposed self-supervised detection for each smartphone model. Training and testing data are collected from the same smartphone model but with different driving traces, i.e., train/test data-splitting is based on driving trace.}
\label{fig:trupos_3.2}
\end{figure}

Next, we conduct self-supervised federated detection within devices of the same smartphone model name but using different traces for training and testing. Compared to Fig.~\ref{fig:trupos_2.2}, which is based on \ac{iid} data where the same traces are used for both training and testing, Fig.~\ref{fig:trupos_3.2} shows a similar performance. However, we observe that their true positive rate is a bit lower than Fig.~\ref{fig:trupos_2.2}. Additionally, detection performance varies across different smartphone models: Phone 1\&4 achieves much higher accuracy than Phone 2\&6. This may be attributed to the larger training set available for Phone 1\&4 ($9899 + 10626$ samples) compared to Phone 2\&6 ($6238 + 6734$ samples). A larger dataset generally improves accuracy, and the transition from an \ac{iid} setting (Fig.~\ref{fig:trupos_2.2}) to a non-\ac{iid} setting (Fig.~\ref{fig:trupos_3.2}) also introduces additional generalization challenges. 

\begin{figure}
\begin{centering}
\includegraphics[width=\columnwidth]{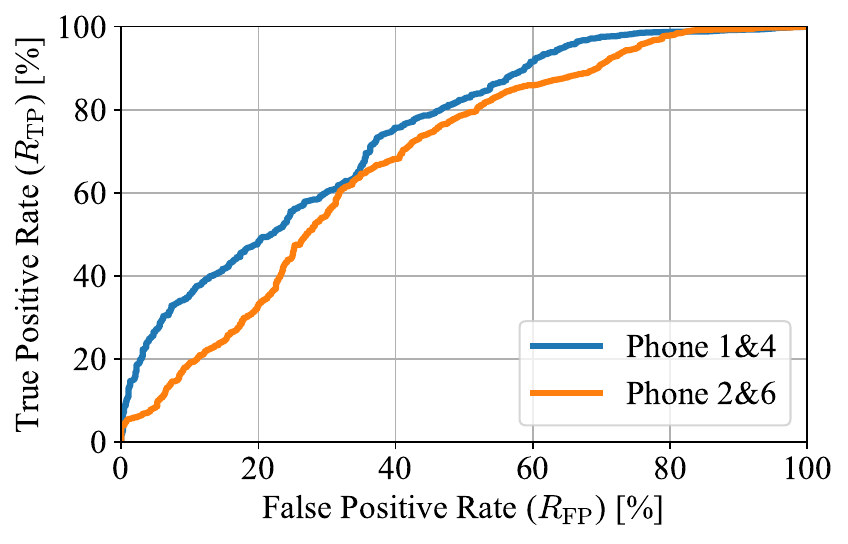}
\par\end{centering}
\caption{\ac{roc} curves of the proposed self-supervised detection for different smartphone models. Training and testing data are collected from different smartphone models and driving traces. Specifically, the curve for Phone 1\&4 is trained on Google Pixel 8 and tested on other phones. Within each model, the train/test split is further based on driving traces.}
\label{fig:trupos_3.3}
\end{figure}

Furthermore, similar to Fig.~\ref{fig:trupos_2.3}, we extend our evaluation by training the detector on devices of one smartphone model using traces of the training set, then testing it on a different smartphone model with different traces. The results are presented in Fig.~\ref{fig:trupos_3.3}. Fig.~\ref{fig:trupos_2.3} and Fig.~\ref{fig:trupos_3.3} show a similar performance, as both involve non-\ac{iid} training and testing settings. The key difference is that Fig.~\ref{fig:trupos_3.3} needs a more comprehensive generalization, since it introduces the train-test split for both the smartphone model and the driving traces. 

\subsection{Discussion}
Our detection method is self-supervised and federatively trained by multiple devices that provide heterogeneous traces and opportunistic data. Unlike existing deep learning-based detections, the proposed method does not require manual annotation for training data, meaning that its training process is self-supervised by the labels generated from a traditional detection, PDS \cite{LiuPap:C23}. Interestingly, while PDS provides training labels for the proposed federated detection, the proposed method outperforms PDS in accuracy. Additionally, without any loss of privacy, each device transmits only model parameters instead of position-related data. 

The proposed federated detection can be generalized to different situations, and we show the generalization of the detection model through a comprehensive evaluation. It evaluates on different devices, smartphone models, and driving traces. The results demonstrate that the method performs robustly on \ac{iid} and non-\ac{iid} data. 

\subsection{Limitations and Roadmap}
We observe that the detection performance on Pixel 4 XL (Phone 2 and Phone 6) is lower than that on other smartphones. Based on feature engineering, this may be because Pixel 4 XL has a significantly different distribution of signal power from satellites than other smartphones. Furthermore, the dataset size, including network position data, for the Pixel 4 XL is much smaller than that of the Pixel 8. 

Our next step is to implement this self-supervised federated detection in a real-world wireless communication environment. Although the current evaluation is based on a real-world dataset, the federated learning algorithm has not yet been deployed on actual mobile platforms with live communication. Communication will introduce network delay and loss, which may affect our detection. Furthermore, we currently consider all mobile platforms to be benign and honestly contribute model parameters. If attackers participate in model aggregation, we need to introduce more practical defense mechanisms, supported by theoretical analysis, to defend against attacks. 

\section{Conclusion}
\label{sec:conclu}
In this paper, we present self-supervised federated \ac{gnss} spoofing detection leveraging opportunistic data that enables mobile platforms to share knowledge about \ac{gnss} spoofing without leaking position privacy. Furthermore, it achieves better privacy preservation and detection performance than existing position-based and deep learning-based methods due to the expected benefits of self-supervised and federated \ac{gnss} spoofing detection. Labels are generated by PDS \cite{LiuPap:C23}, which are used to train local \ac{lstm} models immediately. The \ac{lstm} input features include estimated position deviations and features derived from \ac{gnss} signals. Local model parameters are uploaded to a server, not the actual measurements, and then the server performs model aggregation and quality control. Our evaluation using a real-world dataset from Jammertest 2024 shows performance improvements and exhibits a good generalization across different devices, smartphone models, and driving traces. 

\bibliographystyle{IEEEtran}
\bibliography{reference/references}

\begin{thebibliography}{10}
\providecommand{\url}[1]{#1}
\csname url@samestyle\endcsname
\providecommand{\newblock}{\relax}
\providecommand{\bibinfo}[2]{#2}
\providecommand{\BIBentrySTDinterwordspacing}{\spaceskip=0pt\relax}
\providecommand{\BIBentryALTinterwordstretchfactor}{4}
\providecommand{\BIBentryALTinterwordspacing}{\spaceskip=\fontdimen2\font plus
\BIBentryALTinterwordstretchfactor\fontdimen3\font minus
  \fontdimen4\font\relax}
\providecommand{\BIBforeignlanguage}[2]{{%
\expandafter\ifx\csname l@#1\endcsname\relax
\typeout{** WARNING: IEEEtran.bst: No hyphenation pattern has been}%
\typeout{** loaded for the language `#1'. Using the pattern for}%
\typeout{** the default language instead.}%
\else
\language=\csname l@#1\endcsname
\fi
#2}}
\providecommand{\BIBdecl}{\relax}
\BIBdecl

\bibitem{Goo:J22}
\BIBentryALTinterwordspacing
D.~Goodin, ``{GPS} interference caused {FAA} reroute {Texas} air traffic.
  experts stumped,'' \emph{Ars Technica}, 2022. [Online]. Available:
  \url{https://arstechnica.com/information-technology/2022/10/cause-is-unknown...}
\BIBentrySTDinterwordspacing

\bibitem{SheWonCheChe:C20}
J.~Shen, J.~Y. Won, Z.~Chen, and Q.~A. Chen, ``Drift devil: Security
  multi-sensor fusion based localization high-level autonomous driving {GPS}
  spoofing,'' in \emph{Proc. 29th USENIX Security}, virtual event, Aug. 2020.

\bibitem{BinZiwYonLia:J20}
Q.~Bin, C.~Ziwen, X.~Yong, H.~Liang, and S.~Sheng, ``{GPS} spoofing-based time
  synchronisation attack advanced metering infrastructure its protection,''
  \emph{J. Eng.}, vol. 2020, no.~9, pp. 809--815, 2020.

\bibitem{LiuCheYanShu:C21}
S.~Liu, X.~Cheng, H.~Yang, Y.~Shu, X.~Weng, P.~Guo, K.~C. Zeng, G.~Wang, and
  Y.~Yang, ``Stars can tell: Robust method defend {GPS} spoofing attacks using
  off-the-shelf chipset,'' in \emph{Proc. 30th USENIX Security}, virtual event,
  Aug. 2021.

\bibitem{KhaRosLanCha:C14}
S.~Khanafseh, N.~Roshan, S.~Langel, F.-C. Chan, M.~Joerger, and B.~Pervan,
  ``{GPS} spoofing detection using {RAIM} {INS} coupling,'' in \emph{Proc.
  IEEE/ION PLANS}, Monterey, CA, USA, May 2014.

\bibitem{MaaKas:J21}
M.~Maaref and Z.~M. Kassas, ``Autonomous integrity monitoring vehicular
  navigation cellular signals opportunity {IMU},'' \emph{IEEE Trans. Intell.
  Transp. Syst.}, vol.~23, no.~6, pp. 5586--5601, 2021.

\bibitem{RotCheLoWal:J21}
F.~Rothmaier, Y.-H. Chen, S.~Lo, and T.~Walter, ``A framework {GNSS} spoofing
  detection combinations metrics,'' \emph{IEEE Trans. Aerosp. Electron. Syst.},
  vol.~57, no.~6, pp. 3633--3647, 2021.

\bibitem{LiuPap:C23}
W.~Liu and P.~Papadimitratos, ``Probabilistic detection {GNSS} spoofing using
  opportunistic information,'' in \emph{Proc. IEEE/ION PLANS}, Monterey, CA,
  USA, Apr. 2023.

\bibitem{MorDelLoh:J19}
R.~Morales~Ferre, A.~de~la Fuente, and E.~S. Lohan, ``Jammer classification
  {GNSS} bands machine learning algorithms,'' \emph{Sensors}, vol.~19, no.~22,
  p. 4841, 2019.

\bibitem{BorLiWuClo:J24}
P.~Borhani-Darian, H.~Li, P.~Wu, and P.~Closas, ``Detecting {GNSS} spoofing
  using deep learning,'' \emph{EURASIP J. Adv. Signal Process.}, vol. 2024,
  no.~1, p.~14, 2024.

\bibitem{WuCalImbClo:C23}
P.~Wu, H.~Calatrava, T.~Imbiriba, and P.~Closas, ``Jammer classification
  federated learning,'' in \emph{Proc. IEEE/ION PLANS}, Monterey, CA, USA, Apr.
  2023.

\bibitem{DenLuoYao:C24}
M.~Deng, R.~Luo, and Z.~Yao, ``{GNSS} interference signal classification based
  federated learning,'' in \emph{Proc. 100th IEEE VTC}, Washington, DC, USA,
  Oct. 2024.

\bibitem{NguMarMieFer:C19}
T.~D. Nguyen, S.~Marchal, M.~Miettinen, H.~Fereidooni, N.~Asokan, and A.-R.
  Sadeghi, ``{D{\"I}oT}: Federated self-learning anomaly detection system
  {IoT},'' in \emph{Proc. 39th IEEE ICDCS}, Dallas, TX, USA, Jul. 2019.

\bibitem{FerDmiRieMie:C22}
H.~Fereidooni, A.~Dmitrienko, P.~Rieger, M.~Miettinen, A.-R. Sadeghi, and
  F.~Madlener, ``{FedCRI}: Federated mobile cyber-risk intelligence,'' in
  \emph{Proc. NDSS}, San Diego, CA, USA, Apr. 2022.

\bibitem{Jam:J24}
\BIBentryALTinterwordspacing
Jammertest, ``The world's largest open jamming spoofing test,''
  \emph{Jammertest}, 2024. [Online]. Available:
  \url{https://jammertest.no/about-2/}
\BIBentrySTDinterwordspacing

\bibitem{LenSpaPap:C21}
M.~Lenhart, M.~Spanghero, and P.~Papadimitratos, ``Relay/replay attacks {GNSS}
  signals,'' in \emph{Proc. 14th ACM WiSec}, virtual event, Jun. 2021.

\bibitem{KasKhaAbdLee:J22}
Z.~M. Kassas, J.~Khalife, A.~A. Abdallah, and C.~Lee, ``I am not afraid {GPS}
  jammer: Resilient navigation signals opportunity {GPS}-denied environments,''
  \emph{IEEE Aerosp. Electron. Syst. Mag.}, vol.~37, no.~7, pp. 4--19, 2022.

\bibitem{PapJov:C08}
P.~Papadimitratos and A.~Jovanovic, ``{GNSS}-based positioning: Attacks
  countermeasures,'' in \emph{Proc. IEEE Mil. Commun. Conf.}, San Diego, CA,
  USA, Nov. 2008.

\bibitem{SchRadCamFoo:J16}
D.~Schmidt, K.~Radke, S.~Camtepe, E.~Foo, and M.~Ren, ``A survey analysis
  {GNSS} spoofing threat countermeasures,'' \emph{ACM Comput. Surv.}, vol.~48,
  no.~4, pp. 1--31, 2016.

\bibitem{NarRanNou:C19}
S.~Narain, A.~Ranganathan, and G.~Noubir, ``Security {GPS/INS} based on-road
  location tracking systems,'' in \emph{Proc. IEEE S\&P}, San Francisco, CA,
  USA, May 2019.

\bibitem{XuYinLi:J20}
J.~Xu, S.~Ying, and H.~Li, ``{GPS} interference signal recognition based
  machine learning,'' \emph{Mob. Netw. Appl.}, vol.~25, no.~6, pp. 2336--2350,
  2020.

\bibitem{BorLiWuClo:C20}
P.~Borhani-Darian, H.~Li, P.~Wu, and P.~Closas, ``Deep neural network approach
  detect {GNSS} spoofing attacks,'' in \emph{Proc. 33rd ION GNSS+}, virtual
  event, Sep. 2020.

\bibitem{TohMos:C20}
S.~Tohidi and M.~R. Mosavi, ``Effective detection {GNSS} spoofing attack using
  multi-layer perceptron neural network classifier trained {PSO},'' in
  \emph{Proc. 25th CSICC}, Tehran, Iran, Jan. 2020.

\bibitem{KarPalJay:C21}
D.~R. Kartchner, R.~Palmer, and S.~K. Jayaweera, ``Satellite navigation
  anti-spoofing using deep learning receiver network,'' in \emph{Proc. IEEE
  CCAAW}, Cleveland, OH, USA, Jun. 2021.

\bibitem{ElaUjaRuo:C22}
A.~Elango, S.~Ujan, and L.~Ruotsalainen, ``Disruptive {GNSS} signal detection
  classification different power levels using advanced deep-learning
  approach,'' in \emph{Proc. 12th ICL-GNSS}, Tampere, Finland, Jun. 2022.

\bibitem{FenSeoCao:C22}
Z.~Feng, C.~K. Seow, and Q.~Cao, ``{GNSS} anti-spoofing detection based
  gaussian mixture model machine learning,'' in \emph{Proc. 25th IEEE ITSC},
  Macau, China, Oct. 2022.

\bibitem{CalBhaBovGiu:C20}
R.~Calvo-Palomino, A.~Bhattacharya, G.~Bovet, and D.~Giustiniano, ``Short:
  {LSTM}-based {GNSS} spoofing detection using low-cost spectrum sensors,'' in
  \emph{Proc. 21st IEEE WoWMoM}, Cork, Ireland, Aug. 2020.

\bibitem{KaaMakRuoMal:C21}
S.~Kaasalainen, M.~M{\"a}kel{\"a}, L.~Ruotsalainen, T.~Malmivirta, T.~Fordell,
  K.~Hanhij{\"a}rvi, A.~Wallin, T.~Lindvall, and S.~Nikolskiy,
  ``Reason-resilience security geospatial data critical infrastructures,'' in
  \emph{Proc. 11th ICL-GNSS}, Tampere, Finland, Jun. 2021.

\bibitem{McmMooRamHam:C17}
B.~McMahan, E.~Moore, D.~Ramage, S.~Hampson, and B.~A. y~Arcas,
  ``Communication-efficient learning deep networks decentralized data,'' in
  \emph{Proc. 20th AISTATS}, Ft. Lauderdale, FL, USA, Apr. 2017.

\bibitem{HarRaoMatRam:J18}
A.~Hard, K.~Rao, R.~Mathews, S.~Ramaswamy, F.~Beaufays, S.~Augenstein,
  H.~Eichner, C.~Kiddon, and D.~Ramage, ``Federated learning mobile keyboard
  prediction,'' \emph{arXiv preprint arXiv:1811.03604}, 2018.

\bibitem{JalRavBadUch:C21}
D.~Jallepalli, N.~C. Ravikumar, P.~V. Badarinath, S.~Uchil, and M.~A. Suresh,
  ``Federated learning object detection autonomous vehicles,'' in \emph{Proc.
  7th IEEE BigDataService}, Oxford, United Kingdom, Aug. 2021.

\bibitem{SheEdwReiMar:J20}
M.~J. Sheller, B.~Edwards, G.~A. Reina, J.~Martin, S.~Pati, A.~Kotrotsou,
  M.~Milchenko, W.~Xu, D.~Marcus, R.~R. Colen \emph{et~al.}, ``Federated
  learning medicine: Facilitating multi-institutional collaborations without
  sharing patient data,'' \emph{Sci. Rep.}, vol.~10, no.~1, p. 12598, 2020.

\bibitem{RieKraMieDmi:J22}
P.~Rieger, T.~Krau{\ss}, M.~Miettinen, A.~Dmitrienko, and A.-R. Sadeghi,
  ``Close gate: Detecting backdoored models federated learning based
  client-side deep layer output analysis,'' \emph{arXiv preprint
  arXiv:2210.07714}, 2022.

\bibitem{KraDmi:J23}
T.~Krau{\ss} and A.~Dmitrienko, ``Avoid adversarial adaption federated learning
  multi-metric investigations,'' \emph{arXiv preprint arXiv:2306.03600}, 2023.

\bibitem{MorRamPanMon:J22}
Y.~More, P.~Ramachandran, P.~Panda, A.~Mondal, H.~Virk, and D.~Gupta,
  ``{SCOTCH}: Efficient secure computation framework secure aggregation,''
  \emph{arXiv preprint arXiv:2201.07730}, 2022.

\bibitem{ZhaLarPap:J22}
K.~Zhang, E.~G. Larsson, and P.~Papadimitratos, ``Protecting {GNSS} open
  service navigation message authentication distance-decreasing attacks,''
  \emph{IEEE Trans. Aerosp. Electron. Syst.}, vol.~58, no.~2, pp. 1224--1240,
  2022.

\bibitem{LenSpaPap:C22}
M.~Lenhart, M.~Spanghero, and P.~Papadimitratos, ``Distributed mobile message
  level relaying/replaying {GNSS} signals,'' in \emph{Proc. ION ITM}, Long
  Beach, CA, USA, Jan. 2022.

\bibitem{SpaPap:J24}
M.~Spanghero and P.~Papadimitratos, ``Time-based {GNSS} attack detection,''
  \emph{IEEE Trans. Aerosp. Electron. Syst.}, pp. 1--18, 2024.

\end{thebibliography}

\end{document}